# Quantum RNG Integration in an NG-PON2 Transceiver

Nemanja Vokić, Dinka Milovančev, Christoph Pacher, Martin Achleitner,
Hannes Hübel and Bernhard Schrenk

AIT Austrian Institute of Technology, Center for Digital Safety&Security / Security & Communication Technologies, 1210 Vienna, Austria.

**Abstract**

We propose a minimally invasive optical overlay for time-interleaved QRNG operation in NG-PON2 transceivers. We experimentally demonstrate that random numbers can be generated at a 0.5 Gb/s rate and validate the randomness through statistical tests.

## I) Introduction

Random numbers are an essential resource in communication and science. Randomness is required to establish seed keys for cryptographic protocols, both for (classical) public key infrastructure as well as in quantum cryptography. Simulation and random sampling algorithms also require a source of randomness, as do consumer-oriented applications such as gaming or lotteries. Random number generators based on quantum effects (QRNG) offer an information-theoretic provable way to produce an unpredictable bit stream. The technological developments of QRNGs are well advanced with differing physical implementations of the entropy source, such as path superposition [1], vacuum fluctuations [2, 3], photon number statistics [4], or laser phase noise [5, 6].

We have recently demonstrated the multi-purpose use of coherent optics such as balanced receivers or optical I/Q modulators [7] for QRNG, by measurement of the vacuum state through mixing with a local laser field in a balanced homodyne detector configuration. However, coherent component technology is mainly applied to metro and core networks, while direct detection still prevails in short-reach networks by virtue of its cost efficiency.

In this work, we experimentally demonstrate that QRNG functionality can be conducted through the transceiver architecture of an NG-PON2 optical network unit (ONU). For this, we exploit wavelength-stacking in access networks to overlay an optical QRNG within the ONU. As we will prove, the time-sharing of ONU assets enables to simultaneously generate truly random numbers by eroding two out of four NG-PON2 lanes, while the remaining lanes perform data transmission. A QRNG rate of 500 Mb/s is accomplished and validated through statistical tests.

## II) Multi-Purpose ONU Transceivers for Data Transmission and Quantum Random Number Generation

A representative access network where QRNG functionality is integrated at the head- and tail-end is presented in Fig. 1. NG-PON2 ONUs are populating the field and are connected through a power-splitting based optical distribution network (ODN) to the central office (CO), where they connect to an optical line terminal (OLT). While cost-sharing among OLTs allows to provide a dedicated QRNG with high generation rate at the centralized point of presence, ONUs can neither cost- nor time-share a QRNG. Instead, a cost-effective approach is required to integrate a QRNG at every ONU. In this paper, we exploit multi-purpose photonics to realize such a QRNG, without blocking the entire NG-PON2 traffic of the ONU during random number generation. Figure 2a shows the proposed ONU architecture. Down- and upstream transmission occurs in two wavebands, termed red and blue (R/B), respectively. The aggregation of four channels is conducted through arrayed waveguide gratings (AWG), which can be further conceived as cyclic 4×4 (de)multiplexers. The channel wavelengths $\lambda_i$ are associated to the optical frequencies $\nu_i = \nu_s - 2i\ \Delta\nu$, where $\Delta\nu$ is the channel spacing, e.g., 100 GHz and $\nu_s$ an arbitrary start frequency. The free spectral range for cyclic AWGs can be chosen according to FSR = 16 $\Delta\nu$.

The QRNG overlay capitalizes on this wavelength-stacked ONU architecture and on the wavelength-specific transmission of the AWG demultiplexers. For example, by dedicating the channel $\nu_1$ of the ONU transmitter to the QRNG and by re-tuning it by $\Delta\nu$ to $\nu_0$ through adjustment of temperature or current, this channel is not relayed to the ODN anymore but folded back to the ONU receiver. In case that $\nu_0$ is tuned in a way that is falls between two red channels of the waveband-cyclic AWG at the receiving ONU branch, it can be balanced between channels $\nu_{11}$ and $\nu_{13}$ (Fig. 2a) at the cost of some excess insertion loss at the transmission cross-point at these AWG channels. This resembles the optical feed of a lossy 50/50 split and two broadband photodiodes in a balanced homodyne QRNG [2] that leverages the resources of one upstream and two downstream channels of an NG-PON2 ONU. In the present work, this AWG-based 50/50 split showed a 1.9 dB excess loss with respect to a 50/50 splitter.

The QRNG implementation necessitates only minor modifications of the ONU optics, such as an AWG foldback. It requires the inclusion of a low-noise differential transimpedance amplifier (LN-TIA) for two receiver lanes in addition to the broadband (BB) TIAs, and the tunability of one upstream emitter by ~2 nm. The NG-PON2 traffic can be serviced on two out of four lanes while random numbers are generated. Moreover, the QRNG overlay is not continuously operated as the random numbers can be buffered for later use. It is also possible to simplify the tandem-AWG configuration with preceding waveband split towards a single fold-back AWG with higher port.

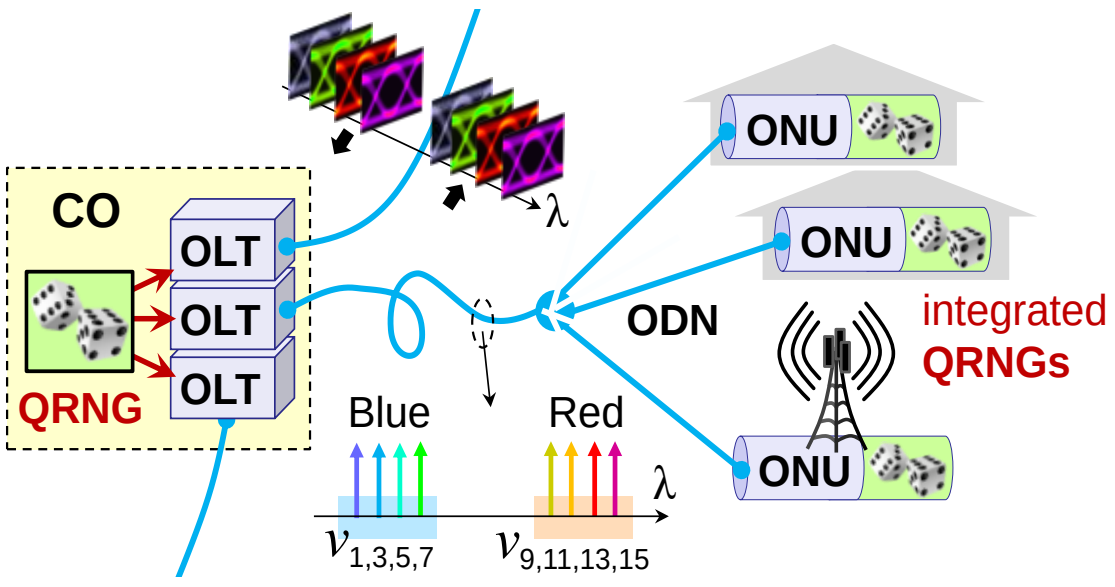

Fig. 1. Seamless PON integration of QRNG functions in ONUs by means of multi-purpose NG-PON2 transceivers.

## III) Experimental Proof for QRNG in NG-PON2 Optical Network Unit

The experimental setup is reported in Fig. 2b. Since N×N AWGs were not available for the ONU, 1×N AWGs have been instead employed in a drop-and-reinject scheme, which necessitates an add-drop (A/D) filter and a 50/50 coupler. The upstream wavelengths corresponding to channels $\nu_{1,3,5,7}$ span from 1550.5 to 1552.7 nm and those for the downstream at $\nu_{9,11,13,15}$ from 1554.3 to 1556.7 nm. The balanced split for $\nu_0$ is accomplished at 1554.85 nm. Figure 3a presents the spectral transmission functions of R/B, A/D and AWG filters together with the wavelength allocation. The downstream channels are modulated at the OLT with 10 Gb/s data signals and transmitted to the ONU over a 13.2 km long single-mode fiber span. The two channels at $\nu_{13,15}$ were routed over an optical switch (SWI) in order to blank these wavelengths during QRNG operation, for which $\nu_1$ is tuned to $\nu_0$ and then overlaps with these two downstream channels (Fig. 3b). After balanced detection at these two receiver branches with the LN-TIA, the signal is digitized and fed to a randomness extractor in order to yield uniformly distributed random bits. Data transmission and QRNG operation is performed according to a pre-defined time frame, which featured a slot with a duration of 2.2 seconds for intermittent QRNG operation and random bit buffering. In the present experiment, the slot width was limited by the laser tuning interval and settling time. Fast tunable lasers [8] would be required to scale down to a duty cycle as known from PON standards. A frame generator synchronizes the downstream blanking through the SWI at the OLT ($\nu_{13,15}$) and the wavelength control of the upstream laser ($\nu_1 \rightarrow \nu_0$) for providing the continuous-wave optical seed at the ONU-side QRNG. A representative transition between operating modes within such a frame is shown in Fig. 3b, for which a settling time of 51 µs can be noticed at the start of the QRNG slot.

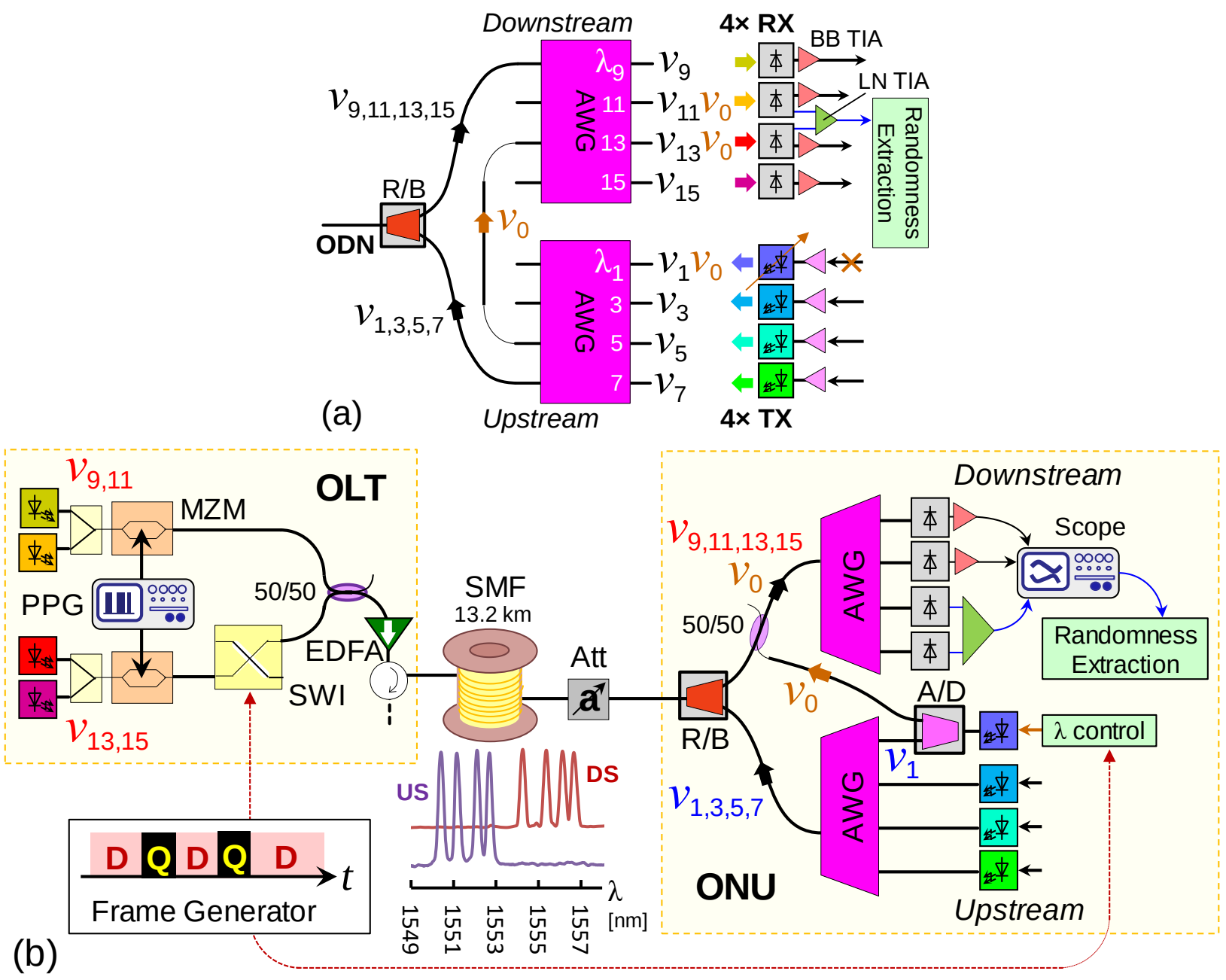

Fig. 2. (a) NG-PON2 ONU architecture for data transmission and QRNG. (b) Experimental setup for QRNG and data transmission in a PON.

## IV) Results and Discussion

Figure 3c reports the accomplished quantum-to-classical noise clearance ξ as function of the reception bandwidth for a seed laser power of 10 dBm at $\nu_0$. The clearance is 14.8 dB at 200 MHz and reduces to 7.5 dB at 600 MHz. This area between the noise spectra for dark and lit balanced receiver with LN-TIA can be exploited for the homodyne QRNG. It shall be noted that this high clearance would not be possible for detectors using the BB-TIA since the electrical background in Fig. 3c is mainly determined by the TIA performance. Further, saturation effects will likely apply for single-ended detectors due to the relatively high optical power at $\nu_0$.

The output signal of the balanced detector was then acquired by a real-time scope and thus digitized with 8-bit resolution for the purpose of randomness extraction based on universal Toeplitz hashing [9]. This step ensures a uniformly randomness among the sampled bits, which in their raw form are Gaussian distributed, show correlations due to the non-ideal transmission function of the balanced detector and the LN-TIA, and crosstalk due to either optical or, more likely, electrical coupling, including EMI coupling that leads to a hidden pattern. The raw samples have been input to a seeded strong randomness extraction algorithm, which aims at improving the entropy per bit. It does so by hashing the random data of the input sample stream using an independent random seed [9].

The randomness extraction has been performed through off-line DSP. We used a rough, non-ITS estimate for the lower bound on min-entropy of 1/4 bit per acquired sample for the purpose of technical demonstration, leaving the accurate estimate for a detailed model of the source. The resulting random bit stream has been examined with the NIST SP800-22-rev1a randomness test suite and the results are shown in Fig. 4. Successful tests require a pass rate above 0.98, with a P-value above 0.01, which is the case for all tests performed on the extracted bits. With the given bandwidth, clearance and entropy after randomness extraction, a generation rate of 0.5 Gb/s appears to be feasible.

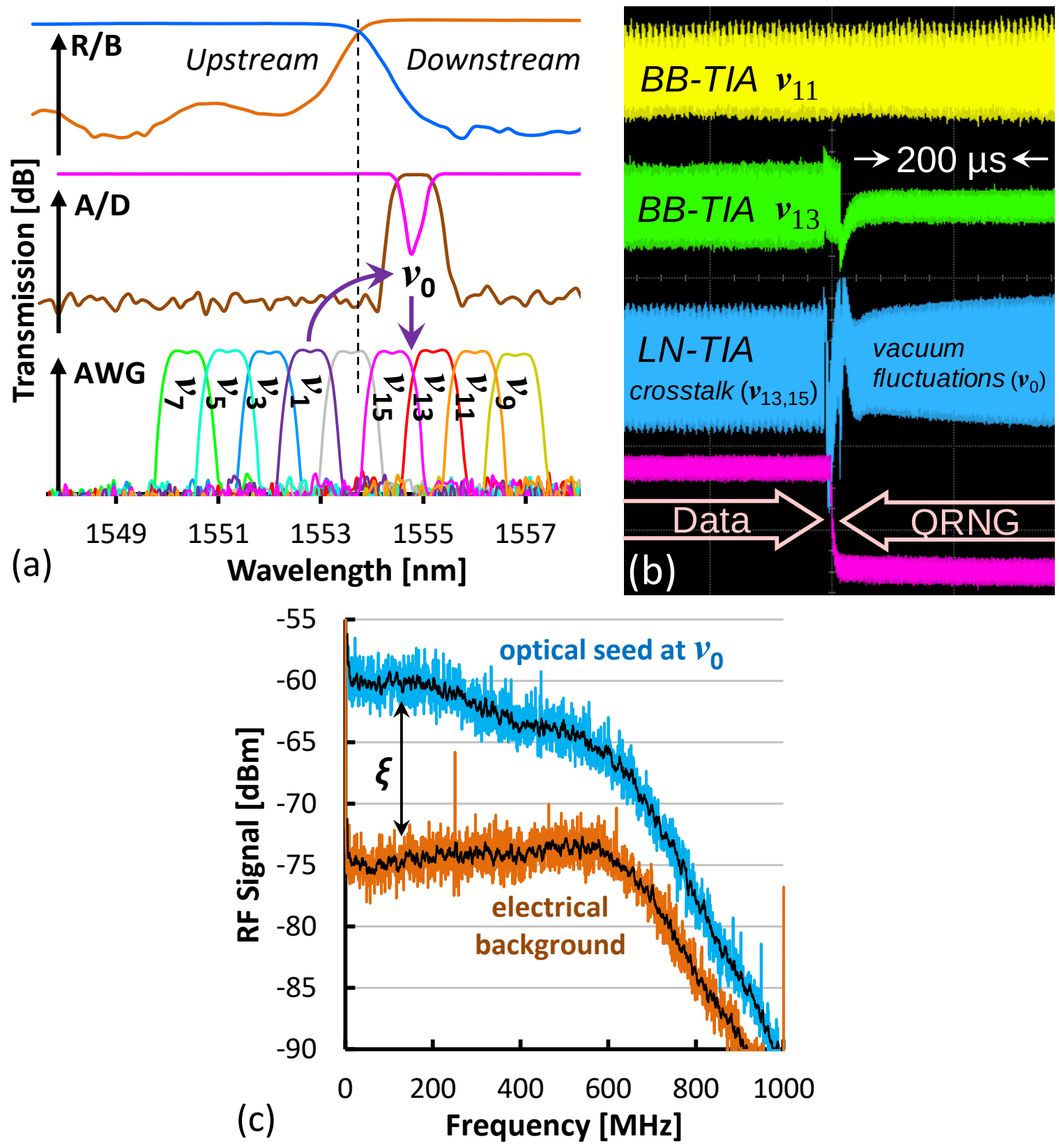

Fig. 3. (a) Filter and AWG transmission spectra in NG-PON2 transceiver.
(b) Time-interleaved QRNG operation. (c) Clearance spectrum.

| | | | | | | | | | | | | | | | |
|---|---|---|---|---|---|---|---|---|---|---|---|---|---|---|---|
| 34 | 3E | FD | 04 | B5 | B2 | B1 | CD | 81 | CE | 45 | 1C | A2 | E2 | 8C | BA |
| 61 | 4E | AC | 9D | 93 | 37 | E4 | 43 | BB | 4D | F6 | DA | 73 | E0 | E4 | 75 |
| 6D | A5 | FC | 49 | FA | 78 | 93 | F3 | 5E | FE | 68 | A2 | 8B | C8 | 34 | F7 |
| 87 | 06 | 19 | D6 | 1E | 7D | B2 | 3C | FE | DB | 95 | F1 | DE | 91 | 26 | 6A |
| 3D | 48 | 3C | D5 | 42 | 73 | 3A | F3 | 71 | A8 | E1 | C4 | 65 | 0A | 61 | 5E |
| 66 | 3E | 8C | 0B | 41 | 2A | FF | 8D | EC | 2B | 1C | F9 | 26 | 53 | 3A | 2D |
| C3 | 62 | 98 | 47 | 2F | 87 | 9B | 05 | 77 | 96 | 97 | AA | 1E | B2 | 0E | 58 |
| 3E | D4 | C2 | A6 | DA | 0B | EB | 88 | 9B | 49 | 9B | E5 | AB | 08 | 45 | 2E |

Tab. 1. 256 bytes of the generated random bit stream.

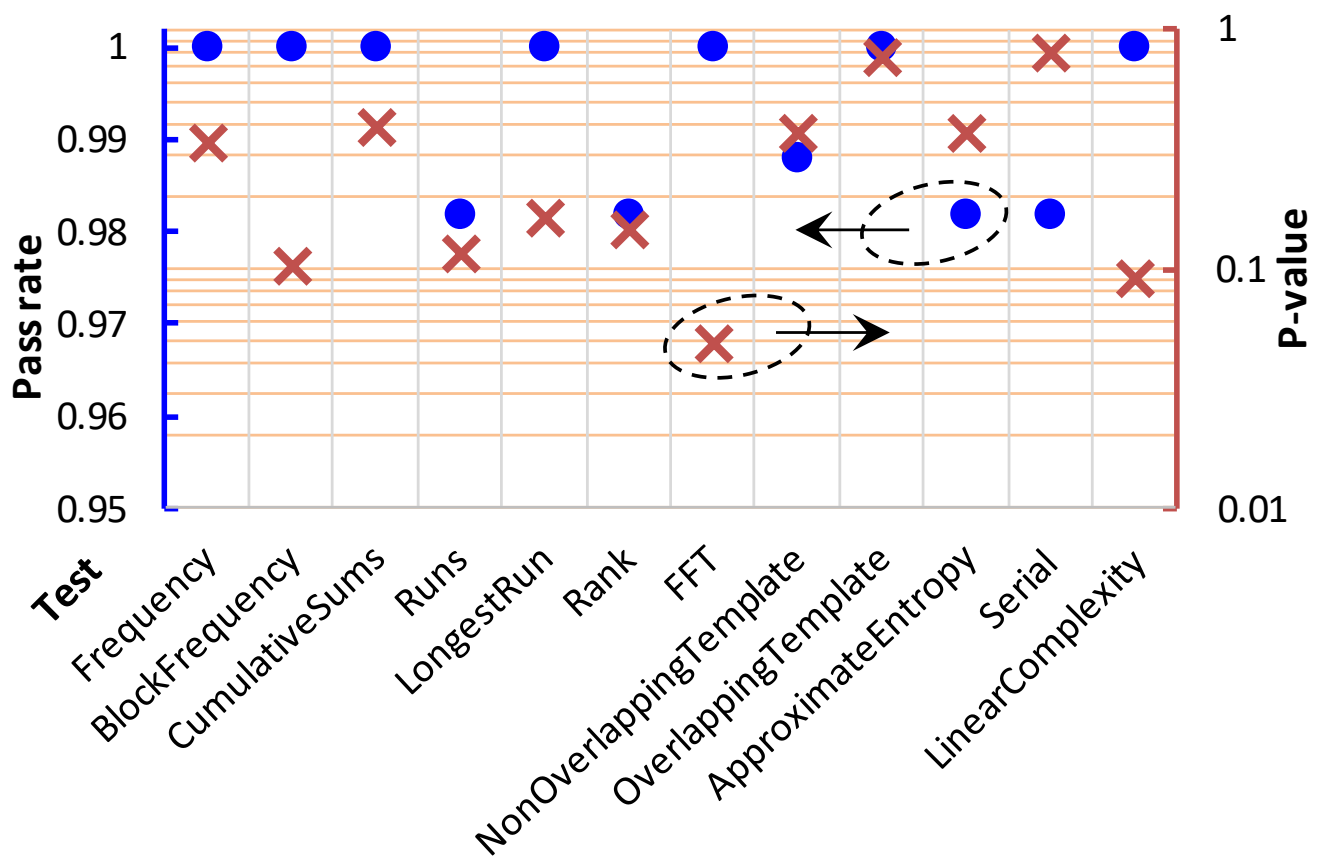

Fig. 4. Statistical tests performed on the extracted random bit stream.

## V) Conclusions

We have experimentally demonstrated the time-interleaved generation of random numbers through a homodyne QRNG overlay in a wavelength-stacked detector architecture, as it applies for NG-PON2 transceivers. The accomplished QRNG rate corresponds to 1.95×10$^{6}$ of generated 256-bit AES keys per second. Eroding this rate for the purpose of periodic key exchange and considering the widely agreed low attack success probability of ~2$^{-50}$ for updating AES keys after transmission of every 64 GByte of data, the obtained QRNG rate is more than sufficient to address the needs in access networks. It can therefore be facilitated with a low duty cycle for QRNG operation.

**VII) Acknowledgement**

This work has received funding from the EU Horizon 2020 programme under grant agreement No 820474 and by the ERC under the grant agreement No 804769.